\documentclass[a4paper]{article}

\usepackage{xspace}
\usepackage[all]{xy}
\usepackage{amssymb}
\usepackage{amsfonts}
\usepackage{amsmath}
\usepackage[latin1]{inputenc}
\usepackage[dvips]{graphicx}
\usepackage{pifont}
\usepackage{tabls}
\usepackage{enumerate}

\newtheorem{theorem}{Theorem}
\newtheorem{corollary}[theorem]{Corollary}

\newtheorem{proposition}[theorem]{Proposition}

\newtheorem{example}{Example}

\makeatletter
\def\@yproof[#1]{\@proof{ #1}}
\def\@proof#1{\noindent\emph{Proof#1.~}}
\def\@ysproof[#1]{\@sproof{ #1}}
\def\@sproof#1{\noindent\emph{Sketch of the proof#1.~}}
\newenvironment{proof}{\@ifnextchar[{\@yproof}{\@proof{}}}
                     {\rule{0pt}{0pt}\hfill \ding{111}\vspace{2ex}}
\newenvironment{sproof}{\@ifnextchar[{\@ysproof}{\@sproof{}}}
                     {\rule{0pt}{0pt}\hfill \ding{111}\vspace{2ex}}
\makeatother
\newcommand{\ie}{\emph{i.e.}\@\xspace}
\newcommand{\wrt}{\emph{w.r.t.}\@\xspace}

\newcommand{\ignore}[1]{}
\newcommand{\Z}{\mathbb{Z}}
\newcommand{\Zt}{\widetilde{\mathbb{Z}}}
\newcommand{\N}{\mathbb{N}}
\newcommand{\card}[1]{\left|#1\right|}
\newcommand{\A}{\ensuremath{\mathcal{A}}\xspace}
\newcommand{\SPM}{\ensuremath{\mathcal{S}}\xspace}
\newcommand{\lSPM}{\lambda_{\mathcal{S}}}
\newcommand{\fSPM}{f_{\mathcal{S}}}
\newcommand{\SPMi}{\ensuremath{\mathcal{S}^{r}}\xspace}
\newcommand{\lSPMi}{\lambda_{\mathcal{S}^{r}}}
\newcommand{\fSPMi}{f_{\mathcal{S}^{r}}}
\newcommand{\LA}{\ensuremath{\mathcal{L}}\xspace}
\newcommand{\lLA}{\lambda_{\mathcal{L}}}
\newcommand{\fLA}{f_{\mathcal{L}}}
\newcommand{\X}{\ensuremath{\mathcal{X}}\xspace}
\newcommand{\lX}{\lambda_{\mathcal{X}}}
\newcommand{\fX}{f_{\mathcal{X}}}
\newcommand{\Y}{\ensuremath{\mathcal{Y}}\xspace}
\newcommand{\lY}{\lambda_{\mathcal{Y}}}
\newcommand{\fY}{f_{\mathcal{Y}}}
\renewcommand{\P}{\ensuremath{\mathfrak{P}}\xspace}
\newcommand{\E}{\ensuremath{\mathfrak{E}}\xspace}
\newcommand{\F}{\ensuremath{\mathfrak{F}}\xspace}
\newcommand{\C}{\ensuremath{\mathfrak{C}}\xspace}
\newcommand{\U}{\ensuremath{\mathfrak{U}}\xspace}
\let\cfg\C
\let\cfgf\C
\let\cfgp\C
\title{Basic properties for sand automata}
\author{J. Cervelle\footnote{Institut Gaspard Monge, Université de
Marne-la-Vallée, \textsc{France},
email: \texttt{julien.cervelle@univ-mlv.fr}}
\and E. Formenti$^{\dag}$ \and B. Masson\footnote{Université de
Nice-Sophia Antipolis, Laboratoire I3S, 2000, Route des lucioles,
Sophia Antipolis, \textsc{France}, email:
\{\texttt{enrico.formenti,benoit.masson}\}\texttt{@I3S.unice.fr}}
}
\date{\today}
\begin{document}

\maketitle
\begin{abstract}
  We prove several results about the relations between injectivity and
  surjectivity for sand automata. Moreover, we begin the exploration
  of the dynamical behavior of sand automata proving that the property
  of nilpotency is undecidable. We believe that the proof technique
  used for this last result might reveal useful for many other results
  in this context.
\end{abstract}

\section{Introduction}
Self-organized criticality (SOC) is a notion which tries to explain
the peculiar behavior of many natural and physical phenomena.
These systems evolve, according to some law, to a ``critical state''.
Any perturbation, no matter how small, of the critical state generate
a deep spontaneous re-organization of the system. Thereafter, the
system evolves to another critical state and so on.

Example of SOC systems are: sandpiles, snow avalanches, star clusters
in the outer space, earthquakes, forest fires, load balance in operating 
systems~\cite{bak97,bak89,bak90,BTW,subramanian94}.

Sandpiles models are a paradigmatic formal model for SOC systems
\cite{GK,GMP}.
In~\cite{CF}, the authors introduced sand automata as a generalization
of sandpiles models and transposed them in the setting of discrete
dynamical systems. A key-point of~\cite{CF} was to introduce a
suitable topology and study the dynamical behavior of sand automata
\wrt this new topology. This resulted in a fundamental
representation theorem similar to the well-known Hedlund's theorem for
cellular automata~\cite{CF,hedlund69}. 

This paper continues the study of sand automata starting from basic
set properties like injectivity and surjectivity.  The decidability of
those two last properties is still an open question. In order to
simplify the decision problem we study the relations between basic set
properties.  We prove that many relations between set properties that
are true in cellular automata are no more true in the context of sand
automata. This allows to conclude that sand automata are a completely
new model and not a peculiar ``sub-model'' of cellular generalized
automata as it might seem at a first glance.

In particular, we show that injective sand automata are not
necessarily reversible but they might have a right inverse automaton
which is not a left inverse. This is a completely new situation
\wrt cellular automata which we think is worthwhile future studies.

Understanding the dynamical behavior of sand automata is in general
very difficult.  Hence we started from very ``simple'' behavior:
nilpotency.  Roughly speaking, a sand automaton is \emph{nilpotent}
if from any starting point it reaches a constant configuration after a
finite number of iterations. We have proved
(Theorem~\ref{th:nilpotency}) that the problem of establishing if a
given automaton is nilpotent is undecidable (when considering spatial
periodic or finite configurations).
 
We believe that the proof technique developed for
Theorem~\ref{th:nilpotency} might be used for proving many other
similar results.

The paper is structured as follows. The next section introduces the
topology on sandpiles and related known results. Section~\ref{sec:sanda}
recalls the definition of sand automata and their representation
theorem. Very interesting and useful examples of sand automata are
presented in Section~\ref{sec:examples}. The main results are in
Sections~\ref{sec:basicsetp} and~\ref{sec:nil}. In Section~\ref{sec:conclu}
we draw our conclusions.
\smallskip

Remark that, due to lack of space, some results have no proof.
Their proofs can be found in the appendix.
 
%
%
%
%
\section{The topology on sandpiles}\label{sec:topo}
A \emph{configuration} represents a set of sand grains, organized in
piles and distributed all over a $d$-dimensional grid. With every point
of the grid $\Z^d$ is associated with the \emph{number of grains} \ie
an element of $\Zt=\Z\cup\{-\infty, +\infty\}$. The value $-\infty$
represents a \emph{sink} and $+\infty$ a \emph{source} of sand grains. 
Hence a configuration is an element of $\Zt^{\Z^d}$. We
will denote by $c_{i_1, \ldots, i_d}$ or $c_i$ the number of grains in
the column of $c$ indexed by the vector $i=(i_1, \ldots,
i_d)$. Finally, for $i\in\Zt$, $\E_i$ is the set of configurations
with the value $i$ at position $(0, \ldots, 0)$. Denote \cfg the set
of all configurations.

A configuration $c$ is \emph{finite} if $\exists k\in\N$ such that for
any vector $i \in \Z^d$, $|i| \geq k$, $c_i = 0$ (remark that we
denote by $|\cdot|$ the infinite norm). The set of finite
configurations will be noted \cfgf. A configuration $c$ is
\emph{periodic} if there is a vector $p \in \Z^d$ such that for any
vector $i \in \Z^d$ and any integer $t \in \Z$, $c_i = c_{i+tp}$; \cfgp
denotes the set of periodic configurations. 
\smallskip

In the remainder of the section, definitions are only given for dimension $1$.
The generalization to higher dimensions is straightforward.
\smallskip

In~\cite{CF}, the authors equipped \cfg with a metric topology defined
in two steps. First, one fixes a reference point (for example the
column of index $0$); then the metric is designed in such a way that
two configurations are at small distance if the have ``the same''
number of grains in a (finite) neighborhood of the reference point. Of
course, one should make more precise the meaning of the sentence
``have the same grains content''. The differences in the number of grains
is quantified by a \emph{measuring device} of size $l\in\N$ 
and reference height $m\in\Z$

$$\beta_l^m(n)=\left\{\begin{array}{c@{\hspace{5mm}}l}
    +\infty & \textrm{if $n>m+l$}\enspace,\\
    -\infty & \textrm{if $n<m-l$}\enspace,\\
    n-m & \textrm{otherwise}.
\end{array}\right.$$

If the difference (in the number grains) between two columns is too
high (resp. too low), then it is declared to be $+\infty$
(resp. $-\infty$).

For any configuration $c \in \Zt^\Z$, $l \in \N$, $l \ne 0$ and $i
\in \Z$, define the following sequence of differences:
$$d_l^i(c)=\left\{\begin{array}{l@{\hspace{5mm}}l}
    (\beta_l^{c_i}(c_{i-l}), \ldots, \beta_l^{c_i}(c_{i-1}),
    \beta_l^{c_i}(c_{i+1}), \ldots, \beta_l^{c_i}(c_{i+l})) &
    \textrm{if $|c_i| \ne \infty$}\enspace,\\[1ex]
    (\beta_l^{0}(c_{i-l}), \ldots, \beta_l^{0}(c_{i-1}),
    \beta_l^{0}(c_{i+1}), \ldots, \beta_l^{0}(c_{i+l})) &
    \textrm{if $c_i = \pm\infty$}\enspace.\\
\end{array}\right.$$

For $l=0$, define $d_0^i(c) = (c_i)$. Finally, the distance between
two configurations $x$ and $y$ is defined as follows: $d(x, y) =
2^{-l}$, where $l$ is the smallest integer such that $d_l^0(x) \ne
d_l^0(y)$.

From now on, \cfg is equipped with the metric topology induced by $d$.
The following propositions prove that the structure of the topology on
\cfg is rich enough to justify the study of dynamical systems on it.

\begin{proposition}[\cite{CF}]
The space \cfg is perfect (\ie it has no isolated point)
and locally compact (\ie for any point $x$ there is a
neighborhood of $x$ whose closure is compact).
\end{proposition}

\begin{proposition}[\cite{CF}]
The space \cfg is totally disconnected (\ie for any points $x, y$
there are two open sets $U$ and $V$ such as $x \in U$, $y \in V$, $U
\cap V = \emptyset$ and $U \cup V = \cfg$).
\end{proposition}

\begin{proposition}[\cite{CF}]
  For any $i\in\Zt$, the set $\E_i$ is compact.
\end{proposition}

The following result completes the characterization
of the topological structure of \cfg.
\begin{proposition}\label{prop:completude}
  The space \cfg is complete.
\end{proposition}

%
\section{Sand automata}\label{sec:sanda}
A \emph{sand automaton} (SA) is a deterministic automaton acting on
configurations. 
It essentially consists in a local rule which is applied in parallel
to each column of the current configuration. The local rules quantifies
the grain content of a neighborhood of the current column to decide
the amount of grains that this columns gains or loses. 
\smallskip

In the sequel, we give the formal definition of sand automaton in
dimension $1$.  Its generalization to higher dimensions is
straightforward.
\smallskip

Formally, a sand automaton is a structure $\A\equiv\langle r, \lambda \rangle$
where $\lambda : \widetilde{[\![-r, r]\!]}^{2r} \rightarrow [\![-r,
r]\!]$ is the local rule and $r$ is the ``size'' (sometimes also called
the \emph{radius}) of the measuring device. The \emph{global function}
$f_{\A}:\cfg \rightarrow \cfg$ of $\A$ is defined as follows
$$\forall c \in \cfg \forall i \in \Z, \quad f_{\A}(c)_i = \left\{
  \begin{array}{l@{\hspace{5mm}}l}c_i & \textrm{if $c_i =
      \pm\infty$}\enspace,\\
    c_i + \lambda(d_r^i(c)) & \textrm{otherwise}.\end{array}\right.$$
    
In~\cite{CF}, the authors show that sand automata can easily
simulate all sandpile models known in literature and even cellular
automata. They also obtained the following fundamental representation
result; but let us first introduce a few more useful definitions.

We need two special functions: the \emph{shift map}
$\sigma : \cfg \rightarrow \cfg$ defined by $\forall c \in \cfg,
\forall i \in \Z, \quad \sigma(c)_i = c_{i+1}$\enspace; and the
\emph{raising map} $\rho : \cfg \rightarrow \cfg$ defined by $\forall
c \in \cfg \forall i \in \Z, \quad \rho(c)_i = c_{i}+1$.  A function
$f:\widetilde{[\![-r, r]\!]}^{2r} \rightarrow [\![-r, r]\!]$ is
\emph{shift invariant} (resp. \emph{vertical invariant}) if $f \circ
\sigma = \sigma \circ f$ (resp. $f \circ \rho = \rho \circ f$). A
function $f:\widetilde{[\![-r, r]\!]}^{2r} \rightarrow [\![-r, r]\!]$
is \emph{infiniteness conserving} if
$$\forall c \in \cfg \forall i \in \Z, \quad \left\{\begin{array}{lcl}
    f(c)_i = +\infty & \Leftrightarrow & c_i = +\infty\\
    \qquad \textrm{and}\\
    f(c)_i = -\infty & \Leftrightarrow & c_i = -\infty \enspace.
  \end{array} \right.$$

\begin{theorem}[\cite{CF}]
A function $f: \cfg \rightarrow \cfg$ is the global function of a sand
automaton if and only if $f$ is continuous, shift-invariant, 
vertical-invariant and infiniteness conserving.
\end{theorem}

By an abuse of terminology, we will often confuse a sand
automaton $\A\equiv\langle r, \lambda \rangle$ with its global
function $f_{\A}$. For example, we will say that $\A$ is \emph{surjective}
(resp. \emph{injective}) if $f_{\A}$ is surjective (resp. injective).  For
$\mathfrak{U}\subseteq\cfg$, $f_{\A}$ is said to be
$\mathfrak{U}$-surjective (resp. injective) if the restriction of $f$
to $\mathfrak{U}$ is surjective (resp. injective).


\section{Examples}\label{sec:examples}

In this section we introduce a series of worked examples with a
twofold purpose: illustrate basic behavior of sand automata and
constitute a set of counter-examples for later use. Some examples
might seem a bit technical but the underlaying ideas are very useful
in the sequel.


\begin{example}\textbf{The automaton \SPM}.\\
This automaton is to the simulation of SPM in dimension $1$: 
$\SPM = \langle 1, \lSPM \rangle$, where
$$\forall x, y \in \widetilde{[\![-1, 1]\!]}, \quad \lSPM(x, y) = \left\{
  \begin{array}{r@{\hspace{5mm}}l}
    +1 & \textrm{if $x = +\infty$ and $y \ne -\infty$}\enspace,\\
    -1 & \textrm{if $x \ne +\infty$ and $y = -\infty$}\enspace,\\
     0 & \textrm{otherwise.}
\end{array}\right.$$

Remark the basic grain movement
of \SPM: a grain falls to the column on
its right when the height difference is bigger than $2$.  
\end{example}
\ignore{
\begin{figure}[!ht]
  \centering
  \includegraphics{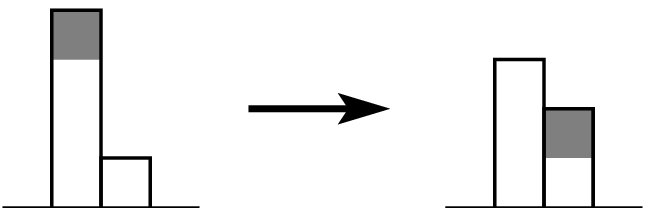}
  \caption{basic behavior of \SPM}
  \label{fig:SPM}
\end{figure}
}

\begin{example}\textbf{The automaton \SPMi}.\\
This automaton is defined similarly to \SPM, but grains climb
the cliffs instead of falling down.
Let $\SPMi = \langle 1, \lSPMi \rangle$ where

$$\forall x, y \in \widetilde{[\![-1, 1]\!]}, \quad \lSPMi(x, y) = \left\{
  \begin{array}{r@{\hspace{5mm}}l}
    -1 & \textrm{if $x = +\infty$ and $y \ne -\infty$}\enspace,\\
    +1 & \textrm{if $x \ne +\infty$ and $y = -\infty$}\enspace,\\
     0 & \textrm{otherwise.}
\end{array}\right.$$
\end{example}
\ignore{
\begin{figure}[!ht]
  \centering \includegraphics{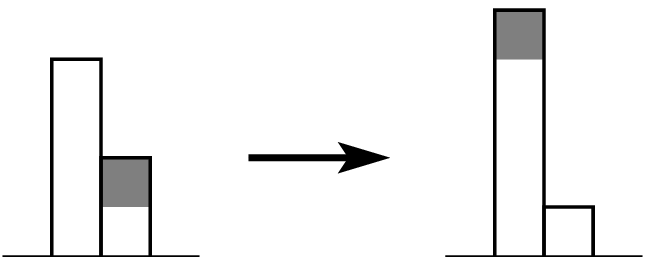}
  \caption{basic behavior of \SPMi}
  \label{fig:SPMi}
\end{figure}
}
\begin{proposition} \label{prop:SPM1}
  The SA \SPM is $\mathfrak{U}$-surjective for $\mathfrak{U}=\C,\F,\P$.
  The SA \SPMi is $\mathfrak{U}$-injective for $\mathfrak{U}=\C,\F,\P$.
\end{proposition}
\begin{proof}
  It is not difficult to see that $\SPM \circ \SPMi = id$, but $\SPMi
  \circ \SPM \ne id$. The first equation implies that \SPM is
  surjective and \SPMi is injective. Moreover, since
  the pre-image by \SPM of a configuration is computed by \SPMi,
  another SA, the pre-image of a finite configuration is
  finite, and periodic if the initial configuration was
  periodic. Hence we have the first part of the thesis. The second
  part is a consequence of the injectivity of \SPMi.
\end{proof}

\begin{proposition} \label{prop:SPM2}
  The SA \SPM is not $\mathfrak{U}$-injective for
  $\mathfrak{U}=\C,\F,\P$.
\end{proposition}

\begin{proposition} \label{prop:SPM3}
  The SA \SPMi is not $\mathfrak{U}$-surjective for
  $\mathfrak{U}=\C,\F,\P$.
\end{proposition}


\begin{example}\textbf{The automaton \LA.}\\
Consider an automaton $\LA = \langle 1, \lLA \rangle$ where 
$$\forall x, y \in \widetilde{[\![-1, 1]\!]}, \quad \lLA(x, y) = \left\{
  \begin{array}{r@{\hspace{5mm}}l}
    -1 & \textrm{if $x < 0$}\enspace,\\
    +1 & \textrm{if $x > 0$}\enspace,\\
     0 & \textrm{otherwise.}
\end{array}\right.$$
Remark the basic behavior of \LA:
each column tries to reach the same height of its left neighbor.
\end{example}
\ignore{
\begin{figure}[!ht]
  \centering
  \includegraphics{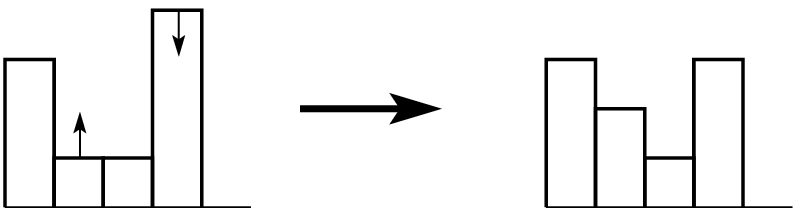}
  \caption{basic behavior of \LA.}
  \label{fig:LA}
\end{figure}
}
\begin{proposition} \label{prop:LA1}
  The SA \LA is not \F-surjective.
\end{proposition}

\begin{proposition} \label{prop:LA2}
  The SA \LA is both \C-surjective and \P-surjective.
\end{proposition}
\begin{proof}
Choose an arbitrary configuration $c$, we are going to build one of
its pre-image $c'$. There is a unique sequence of strictly increasing
indices $(i_n)_{n \in N}$, $N \subset \Z$, such that $\forall i_n \leq
i < i_{n+1},\quad c_i = c_{i_n}$ and $c_{i_n} \ne c_{i_n-1}$ (every
$i_n$ corresponds to a variation in $c$). The idea is to work on these
intervals, amplifying the difference at the border so that an
application of the rule corrects it. Formally, for every $n \in N$,
suppose that $c_{i_n-1} < c_{i_n}$ (if it is not the case then the
symmetrical operations will have to be performed). For every $i_n \leq
i < i_{n+1}$, let $c'_i = c_i + 1$ if $i-i_n$ is even, $c'_i = c_i -
1$ if $i-i_n$ is odd. There are two little subtleties if $N$ is not
bi-infinite. If $n_0 = \inf N$ exists, then let $c'_i = c_i$ for all
$i < n_0$. One could prefer to make it more consistent with the rest
and choose $c'_i = c_i \pm 1$ according to the parity of $i$, it does
not matter much. Second, if $n_1 = \sup N$ exists, then the $\pm$
operation has to be performed forever on the right. Note that it is
why a finite configuration may not have a finite pre-image.

It is not difficult to see that $\fLA(c') = c$. For every $i \in \Z$,
first suppose that there is a $n \in N$ such that $i = i_n$. We have
$\fLA(c')_i = c'_i + \lLA(d^i_1(c'))$. Supposing that $c_{i-1} < c_i$
(again, if it is the opposite then the operations are symmetrical), we
have $c'_i = c_i + 1 > c_{i-1} + 1$, hence $c'_i > c'_{i-1}$ since
$|c_{i-1} - c'_{i-1}| \leq 1$ . So $\lLA(d^i_1(c')) = -1$, and
$\fLA(c')_i = c_i + 1 - 1 = c_i$. Otherwise if $i \ne i_n$ for all $n
\in N$, then by construction we have either:
\begin{itemize}
\item $c'_i = c_i + 1$ and $c'_{i-1} = c_{i-1} - 1 = c_i - 1$, because
  $c$ is constant between the $i_n$'s. Hence $c'_{i-1} = c'_i - 2$,
  and then $\fLA(c')_i = c_i + 1 - 1 = c_i$\enspace;
\item or $c'_i = c_i - 1$ and $c'_{i-1} = c_{i-1} + 1$, the same
  method gives the result.
\end{itemize}

Therefore \LA is surjective. Finally, as the operations we perform on
the configuration are deterministic, a periodic configuration would
have a periodic pre-image (same transformation of the period
everywhere). Hence \LA is also surjective over periodic
configurations.
\end{proof}


The next example is a bit less intuitive since it uses a special
neighborhood: the two nearest left neighbors.

\begin{figure}[!ht]
  \centering
  \includegraphics{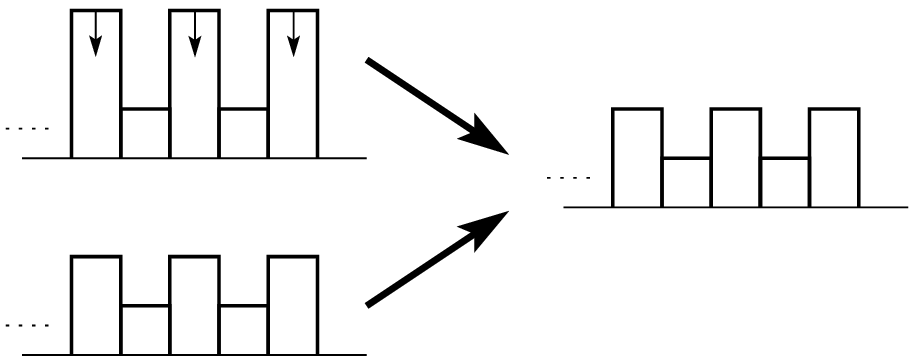}
  \caption{examples of evolution of \X on two different configurations.}
  \label{fig:X}
\end{figure}

\begin{example}\textbf{The automaton \X.}\\
Consider the sand automaton $\X =\langle 2, \lX \rangle$ where 
$$\begin{array}{l@{\hspace{5mm}}l}
  \forall x, y, z \in \widetilde{[\![-2, 2]\!]},
  & \lX(+\infty, x, y, z) = -1\enspace,\\
  & \lX(2, x, y, z) = -1\enspace,\\
  & \lX(1, -1, x, y) = -1\enspace,\\
  & \lX(1, -2, x, y) = -1\enspace,\\
  & \lX(1, -\infty, x, y) = -1\enspace,\\
  & \lX(0, -2, x, y) = -1\enspace,\\
  & \lX(0, -\infty, x, y) = -1\enspace,
\end{array}$$
and any other value gives $0$.  
The behavior of this automaton on two specific sequences is
shown in Figure~\ref{fig:X}. The evolutions of \X on more general
configurations seem quite hard to describe. Anyway, in the
sequel we will need to study its evolutions only on special 
(simple) configurations.
\end{example}

\begin{proposition} \label{prop:X}
  The SA \X is \F-injective but not \C- or \P-injective. 
\end{proposition}

\begin{figure}[!ht]
  \centering
  \includegraphics{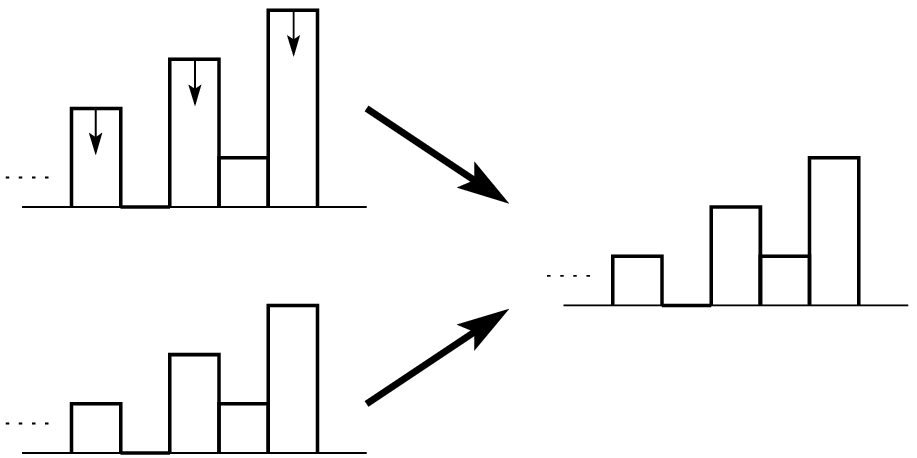}
  \caption{examples of the action of \Y.}
  \label{fig:Y}
\end{figure}

\begin{example}\textbf{The automaton \Y.}\\
  Consider the following SA $\Y = \langle 2, \lY \rangle$,
  where
  $$\begin{array}{l@{\hspace{5mm}}l}
    \forall x, y, z \in \widetilde{[\![-2, 2]\!]},
    & \lY(+\infty, x, y, z) = -1\enspace,\\
    & \lY(2, x, y, z) = -1\enspace,\\
    & \lY(1, x, y, z) = -1\enspace,\\
    & \lY(0, x, y, z) = -1\enspace,\\
    & \lY(-1, -\infty, x, y) = -1\enspace,\\
  \end{array}$$
  and everything else returns $0$. 
  Figure~\ref{fig:Y} shows two meaningful behaviors 
  of the automaton that will be used later.
\end{example}

\begin{proposition} \label{prop:Y}
  The SA \Y is \F- and \P-injective, but not injective.
\end{proposition}
\begin{proof}
  Consider the two configurations $c$ and $c'$ defined as follows
  $$\forall i \in \Z, \qquad \left\{\begin{array}{l}
      c_{2i} = i\\
      c_{2i+1} = i+2\enspace,
    \end{array}\right.\quad
  \left\{\begin{array}{l}
      c'_{2i} = i\\
      c'_{2i+1} = i+3\enspace.
    \end{array}\right.
  $$
  It is not difficult to see that $\fY(c) = \fY(c') = c$
  (see also Figure~\ref{fig:Y}). Hence \Y is not injective.
  In order to show that \Y is injective over finite and periodic
  configurations, we will need an intermediate result: if $c$, $c'$
  are two distinct configurations such that $\fY(c) = \fY(c')$, then
  there are infinitely many differences, of infinitely many different
  values. Practically, we will show that if $c_i > c'_i$ then
  $c_{i-2} > c'_{i-2}$ and $c_{i-2} < c_i$.

  Assume $c_i > c'_i$ for some $i$, and let $f(c)=f(c')$.
  Then, without loss of generality, one can choose $c_i = c'_i +
  1$ (the difference can not be greater than one, because $\lY$ only
  returns $-1$ or $0$). Therefore, a rule which returns $0$ is applied
  to $c'$ at position $i$, which means that $c'_{i-2} \leq c'_i-1$
  (since $\lY(x, -, -, -)$ returns $0$ only if $x \leq -1$).  For the
  same reason, one of the five rules which returns $-1$ is applied to
  $c$ at position $i$, hence $c_{i-2} \geq c_i-1$.  So $c_{i-2} \geq
  c_i-1 = c'_i \geq c'_{i-2}+1 > c'_{i-2}$. The first consequence of
  this inequality is that if there is a difference somewhere, there
  are infinitely many differences, hence \Y is \F-injective. Indeed
  two finite configurations cannot have infinitely many differences,
  so two different finite configurations have a different image.
  
  Moreover, $c_{i-2} = c'_{i-2} + 1$ to ensure $\fY(c) = \fY(c')$. So
  the inequalities above are in reality equalities, in particular
  $c_{i-2} = c_i - 1$. Therefore it holds $\cdots < c_{i-4} < c_{i-2}
  < c_i$, which proves that two different periodic configurations also
  have different images (a periodic configuration contains a finite
  number of different columns, which is contradicted by the above
  inequality).  As a consequence, \Y is \P-injective.
\end{proof}


\section{Basic set properties}\label{sec:basicsetp}

This section concerns the relations between surjectivity and
injectivity, \wrt all, finite and periodic configurations, in the same
way it was done in~\cite{D} for cellular automata. In particular the
relation between \F-injectivity and surjectivity was interesting, as
for cellular automata it can be used to prove undecidability of
surjectivity~\cite{durand94}. Unfortunately, there is not relation
between those two properties in the context of SA (see
Propositions~\ref{prop:SPM1}, \ref{prop:SPM2}, \ref{prop:SPM3}).
In this section we try to analyze these relation deeper hoping
this might help for the proof of the decidability result about
surjectivity or injectivity.



%
%

\begin{proposition} \label{prop:surjf_surj}
  \F-surjectivity implies surjectivity.
\end{proposition}
\begin{proof}
  For any configuration $c$, let $c^0_n$ be such that $\forall
  i\in\Z,\;(c^0_n)_i=c_i$ if $-n\leq i\leq n$ and $(c^0_n)_i=0$
  otherwise. Consider a sand automaton $f$ that is \F-surjective and
  choose an arbitrary configuration $c\in\C$.  For any $n\in\N$, let
  $c_n=f^{-1}(c^0_n)$. The pre-images $c_n$ are
  contained in some set $E_i$ for $i\in I$, with $\card{I}<\infty$. 
  Since $\cup_{i\in I}E_i$ is compact and 
  $(c_n)_{n\in\N}\subset\cup_{i\in I}E_i$, $(c_n)_{n\in\N}$ contains
  a converging sub-sequence $(c_{n_i})_{i\in\N}$. 
  Let $c^*=\lim_{i\to\infty} c_{n_i}$. By contradiction, assume that
  $f(c^*)\ne c$. Then there exists $j\in\Z$ such that
  $f(c^*)_j\ne c_j$ but $f(c_{n_i})_j=c_j$ for $n_i$ big enough.
  \ignore{
    For any configuration $c$ and any dimension $d$, let $c^0_n$ be such
    that $\forall i\in\Z^d,\;(c^0_n)_i=c_i$ if $|i| \leq n$ and
    $(c^0_n)_i=0$ otherwise. Consider a sand automaton $f$ that is
    \F-surjective and choose an arbitrary configuration $c\in\C$.  For
    any $n\in\N$, let $c_n=f^{-1}(c^0_n)$. For any $n$, the pre-images
    $c_n$ are contained in some set $\E_i$ with $i\in I \subset
    [\![c_0-r, c_0+r]\!]$ (as usual, because $\lambda$ returns a value
    between $-r$ and $r$), hence $\card{I}<\infty$.  Since $\cup_{i\in
      I}\E_i$ is compact and $(c_n)_{n\in\N}\subset\cup_{i\in
      I}\E_i$, $(c_n)_{n\in\N}$ contains a converging sub-sequence
    $(c_{n_i})_{i\in\N}$.  Let $c^*=\lim_{i\to\infty} c_{n_i}$. By
    contradiction, assume that $f(c^*)\ne c$. Then there exists $j\in\Z$
    such that $f(c^*)_j\ne c_j$ but $f(c_{n_i})_j=c_j$ for $n_i$ big
    enough.
  }
\end{proof}

  Remark that the result of Proposition~\ref{prop:surjf_surj} is true
  in any dimension but the opposite implication is false (even in
  dimension $1$), since \LA which is surjective but not \F-surjective
  (see Propositions~\ref{prop:LA1} and \ref{prop:LA2}).

\begin{proposition} \label{prop:surjp_surj}
  \P-surjectivity implies \C-surjectivity.
\end{proposition}

\begin{proposition} \label{prop:surj_surjp}
  In dimension $1$, \C-surjectivity implies \P-surjectivity.
\end{proposition}
\begin{proof}
  Let \A be a surjective sand automaton in dimension 1, and $c^0$ a
  periodic configuration of period $p \in \Z$. Let $c$ be a pre-image
  of $c^0$ by \A. We will build a periodic configuration from $c$,
  whose image will be $c^0$. Let $X = \{ (c_{k-r}, \ldots, c_{k+r-1})
  \:|\: \exists \alpha \in \Z, k = \alpha p \}$. Since for every $i
  \in \Z$, $|c_i - c^0_i| \leq r$ (as $\lambda$ returns an element of
  $[\![-r, r]\!]$), and because $c^0$ is $p$-periodic, there are at
  most $2r \cdot (2r+1)$ elements in $X$.
  Let $k_1 = \alpha_1 p$ and $k_2 = \alpha_2 p$, $k_1 < k_2$ such that
  $(c_{k_1-r}, \ldots, c_{k_1+r-1}) = (c_{k_2-r}, \ldots, c_{k_2+r-1})$.
  Let the $(k_2-k_1)$-periodic configuration $c'$ where the period is
  defined by (see Figure~\ref{fig:period} for the construction)
  $c'_{k_1+i} = c_{k_1+i}$ for all $0 \leq i < k_2-k_1$\enspace.
  \begin{figure}[!ht]
    \centering
    \input{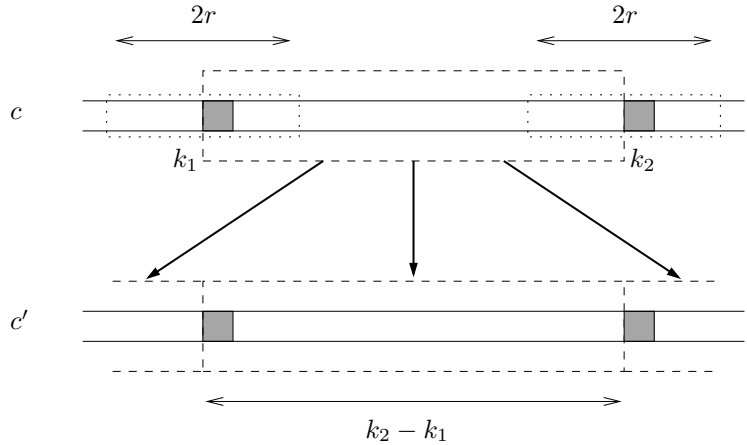}
    \caption{Construction of $c'$ using $c$}
    \label{fig:period}
  \end{figure}
  It is easy to see that $f(c') = c^0$, because for every configuration of the
  period of $c'$, the automaton sees the same neighborhood as for $c$
  (due to the construction of $c'$), so it acts in the same correct
  way. And as $k_2 - k_1$ is a multiple of $p$, each period of $c'$
  coincides with a period of $c^0$, so the image of $c'$ is equal to
  $c$ everywhere: \A is \P-surjective.
\end{proof}

  In dimensions greater than $1$, the above problem is currently open, we have
  no direct proof nor counter-example. The problem is due to the fact
  that in dimension $2$ and above, the size of the perimeter of a ball
  (the $2r$ sequence we used in X for the proof in dimension $1$) is
  linked to the size of the ball. Therefore we can not say that there
  is a finite number of perimeters, and then stick them together to
  build the periodic configuration.

\begin{corollary}\label{cor:surjf-surjp}
  In dimension $1$, \F-surjectivity implies \P-surjectivity.
\end{corollary}

  The question if the above corollary is true in dimension
  $2$ and above is still open and its solution appears to be
  quite difficult.

  Note that the opposite implication of
  Corollary~\ref{cor:surjf-surjp} is false in any dimension, thanks
  to \LA which is \P-surjective but not \F-surjective (see
  Propositions~\ref{prop:LA1} and \ref{prop:LA2}).
\smallskip




  If being injective means a lot, the opposite is not true.
  In fact, because of \X, \F-injectivity does not imply injectivity
  (Proposition~\ref{prop:X}); and \Y shows that \P-injectivity does
  not mean global injectivity (Proposition~\ref{prop:Y}).
The following proposition will complete these results.

\begin{proposition}\label{prop:pinjfinj}
\P-injectivity implies \F-injectivity.
\end{proposition}
\begin{proof}
  This will be proved using the contrapositive. Let \A be an automaton
  not \F-injective. Let $c^1, c^2$ be the two distinct finite
  configurations which lead to the same image $c$. Let $k \in \N$ such
  that for all $i \in \Z^d$, $|i| > k$, $c^1_i = c^2_i = 0$. We are
  going to build two distinct periodic configurations by surrounding
  the non-zero part of $c^1$ and $c^2$ with a crown of zeros, of
  thickness $r$, and repeat this pattern (see Figure~\ref{fig:crown}
  for an illustration in dimension 2).
  
  For $\alpha \in \{1,2\}$, let $d^{\alpha}$ be the
  $(2k+2r+1)$-periodic configuration defined by
  $$\forall i \in \Z^d, |i| \leq k+r, \qquad
  \left\{\begin{array}{l@{\hspace{5mm}}l}
      d^{\alpha}_i = c^{\alpha}_i & \textrm{if $|i| \leq k$}\enspace,\\
      d^{\alpha}_i = 0 & \textrm{if $k < |i| \leq k+r$}\enspace.\\
    \end{array}\right.$$

  \begin{figure}[!ht]
    \centering \input{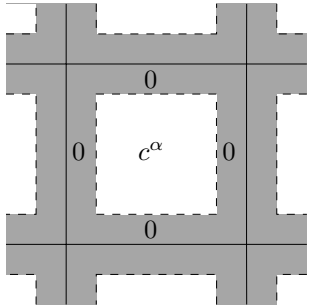}
    \caption{Construction of $d^\alpha$ in dimension 2. White is for
      non-zero values taken in $c^\alpha$, grey is for 0.}
    \label{fig:crown}
  \end{figure}

  We have $f(d^1) = f(d^2)$. For every configuration, we can consider the
  translated configuration whose index is lower in norm than $k+r$ because of
  the periodicity. This configuration reacts as it did in $c^1$ and $c^2$
  because its neighborhood is the same : inside the $k$ ``circle'', it
  is obvious. If it is inside the crown of $0$'s, then the only non-zero
  values it can see are the values located inside the initial pattern.
  So its behavior is equivalent to the one of the point at the border
  of the initial finite configuration, and \A is not \P-injective.
\end{proof}

\begin{figure}[!ht]
   \centering
   $$\xymatrix @!0 @R=1.5cm @C=1cm {
     I \ar@{=>}[rr] \ar@{=>}[dr] && I_{\P} \ar@{=>}[dl] \\
     & I_{\F} } \hspace{2cm}
   \xymatrix @!0 @R=1.5cm @C=1cm {
     S \ar@<3pt>@{=>}[rr]^1 && S_{\P} \ar@<3pt>@{=>}[ll] \\
     & S_{\F} \ar@{=>}[ul] \ar@{=>}[ur]_1 }
   $$
   \caption{relations between basic set properties. $I$ means
   injectivity and $S$ surjectivity. $I_{\U}$ (resp. $S_{\U}$) means
   injectivity (resp. surjectivity) restricted to \U.}
   \label{fig:recap}
\end{figure}
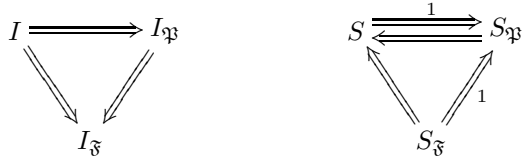

  The opposite implication in Proposition~\ref{prop:pinjfinj} is
  false since \X is \F-injective but not \P-injective
  (Proposition~\ref{prop:X}). Figure~\ref{fig:recap} summarizes the
  relations between basic set properties.

\section{Nilpotency}\label{sec:nil}

Understanding the dynamical behavior of SA seems very difficult. This
is confirmed by the main result of this section: nilpotency is
undecidable for sand automata.

Recall that a SA $f$ is \U-nilpotent if $\forall c\in \U\;\exists
n\in\N$ such that $f^n(c)=\underline{0}$, where $\underline{0}$ is the
configuration in which all columns have no grains.
\medskip

\noindent
\textbf{Problem} \textbf{NIL}(\U)

\noindent
\qquad\textsc{instance}: a SA $\A=\langle\lambda,r\rangle$;

\noindent
\qquad\textsc{question}: is \A \U-nilpotent?

\begin{theorem}\label{th:nilpotency}
Both problems \textbf{NIL}(\P) and \textbf{NIL}(\F) are undecidable.
\end{theorem}
\begin{sproof}
We reduce these problems to the halting problem of a Turing
Machine, simulated by a two counters automaton with finite control.
From such an automaton $A$, we build a SA which is
\F-nilpotent (resp. \P-nilpotent) if
and only if $A$ halts on empty input.
Remark that it is enough to prove the thesis on \F. In fact, from any finite
configuration one can obtain a periodic configuration by repeating
periodically the non-zero pattern surrounded by a suitable border of
zeroes (if necessary).

\begin{figure}[hbtp]
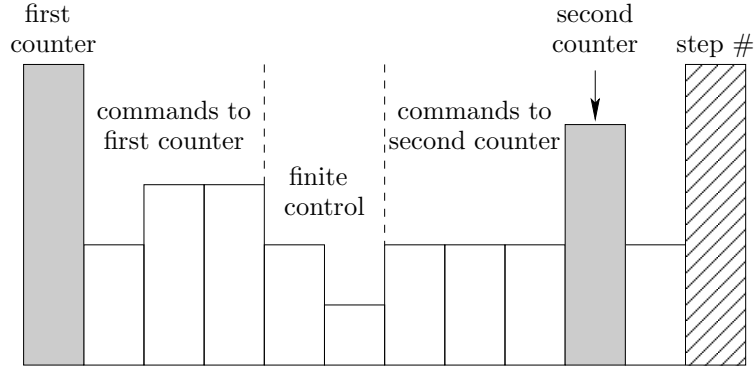

\begin{center}
\input nil2.pstex_t
\end{center}
\caption{simulation of a two counters finite automaton by a SA.}
\label{fig:nil1}
\end{figure}

The simulation is illustrated in Figure~\ref{fig:nil1}. The idea is to
use a certain number of grain stacks for each counter and for the
finite control.  In order to be able to increase or decrease the
height of the stack corresponding to each counter, commands have to be
sent from the finite control to the top of the stacks representing the
counters. This can be done using three stacks: the middle one acts as
a reference, the first one tells if the command is going up to the top
of the counter stack, or down to the finite control. The third one
codes the command: increase or decrease (there is no need for a command
which tests whether the counter is zero or not since the finite
control can verify if stacks are strictly positive or not). Finally,
to make the simulation work properly, one also needs to keep track of
the number of the step in the computation.
To be more precise, each stack has to know who it is, which can be
done using reference stacks as in CA simulation in~\cite{CF}.

Once two counters finite automata can be simulated, it remains to
properly reduce the halting problem to nilpotency. Two points must be
underlined.

First, when a halting state is reached, the configuration must evolve
to $\underline{0}$.  Second, when a configuration is incorrect, and
does not correspond to a real state of the simulated automaton, it
must also annihilate. 

There are two possibilities for a configuration to be incorrect.  It
can just be incorrect with respect to the simulation (for instance,
one of the counter is negative), which is easy to detect. The other
case is when the configuration of the automaton is not reachable
starting from empty input.  These type of errors are a little more
complicated to detect. In fact, at each step the simulator checks,
starting the computation from the very beginning, if the
configuration is correct \ie if the current configuration and the
one obtained from the second simulator are the same. If there is a
difference then the SA evolves to $\underline{0}$; otherwise another
step is performed. The technical details of the construction are left
for the long version of the paper.
\end{sproof}
\section{Conclusions}\label{sec:conclu}
In this paper we have seen that the quest for decidability results for
basic set properties like injectivity and surjectivity is hardened by
the lack of relations between them and their restriction to ``easy''
computable subsets of configurations (such as \P or \F).
This fact can be considered as a first evidence that the study of
dynamical behavior of SA might reveal very difficult.

The second evidence is given by Theorem~\ref{th:nilpotency}. A very
simple dynamical behavior like nilpotency is undecidable. Remark that,
in the case of cellular automata, the undecidability of nilpotency is
a powerful tool for proving the undecidability of many other problems
in cellular automata theory. We think that this property can play a
similar role for sand automata. The authors are currently
investigating this subject.
\nocite{*}

%
\newpage
\appendix
\section{Proofs of remaining results}
\noindent\textbf{Proofs of Section~\ref{sec:topo}.}\\[1ex]
\begin{proof}[of Proposition~\ref{prop:completude}]
Let $(c^n)_{n \in \N}$ be a Cauchy sequence of $\cfg^\N$. There is a
$N \in \N$ such that for all $m, n \geq N$, $d(c^m, c^n) < 1$, in
other words for all $n \geq N$, $c^n_0 = c^N_0$. Every element of the
sequence $(c^{N+n})_{n \in \N}$ is in $\E_{c^N_0}$, which is compact
and hence complete. As this is a Cauchy sequence, it has a limit $c$
in $\E_{c^N_0} \subset \cfg$. $c$ is obviously the limit of the
initial sequence $(c^n)$, which gives the result.
\end{proof}

\noindent\textbf{Proofs of Section~\ref{sec:examples}.}\\[1ex]
\begin{proof}[of Proposition~\ref{prop:SPM2}]
Consider the following finite configurations
$c, c'$ where $c_i = 0$ for $i \in \Z$, $c'_i = 0$ for $i \in \Z
\setminus \{0, 1\}$, $c'_0 = 1$, and $c'_1 = -1$. Clearly, $\fSPM(c) =
\fSPM(c') = c$. Now, consider the periodic configurations $c''$
with $c''_{2i} = 1$ and $c''_{2i+1} = -1$ for every $i \in \Z$, again
$\fSPM(c) = \fSPM(c'') = c$.
\end{proof}

\begin{proof}[of Proposition~\ref{prop:SPM3}]
Consider the following finite configuration $c$, where $c_i=2$ if $i=0$;
$c_i=0$ otherwise. Assume that $c$ has a pre-image $c'$. 
There are only three possibilities for the value of
$c'_0$:
  \begin{description}
  \item[$c'_0 = 3\ :$] then the local rule has to return $-1$, which
    implies that $c'_{-1} \geq 5$. But $\fSPMi(c')_{-1} = 0$, this value
    can not be reached from $5$;
  \item[$c'_0 = 2\ :$] the column is unchanged, which means that
    ($c'_{-1} \leq 3$ or $c'_1 \leq 0$) and ($c'_{-1} \geq 4$ or $c'_1
    \geq 1$). For the same reason as before, $c'_{-1}$ can not be
    greater than $4$, hence $c'_1 \geq 1$. This means that the local
    rule applied at position $1$ returns $-1$, in other words that
    $c'_0 \geq 3$, which contradicts the first hypothesis;
  \item[$c'_0 = 1\ :$] $\lSPMi$ returns $+1$, so $c_1 \leq -1$. Hence
    at position $1$, $\lSPMi$ also returns $+1$. That means, in
    particular, that $c'_2 \leq -3$, which is impossible if one has to
    obtain $\fSPMi(c')_2 = 0$.
\end{description}

We have found a finite configuration with no pre-image, which means that
\SPMi is not surjective both on \C and on \F. To show
that \SPMi is not \P-surjective, one can consider the
configuration $c$ where $c_{4i+1} = 2$ for every $i\in\Z$, and
everywhere else $c_k = 0$. The proof is similar to the previous part,
since the 4 elements of the period act as if the configuration was
finite (radius 1, so they do not ``see'' farther than one column ahead and
one column back).
\end{proof}

\begin{proof}[of Proposition~\ref{prop:LA1}]
  Consider the finite configuration $c$ where $c_i = 2$ if $i=0$
and $c_i=0$ otherwise. By contradiction assume that $c'$ is the
pre-image of $c$ and that $c'\in \cfgf$.
Let $i$ be the greatest integer such that $c'_i \ne
0$. Then since $c'_i \ne 0$ and $c'_{i+1} = 0$, it holds that 
$\fLA(c')_{i+1}=c_{i+1}\ne0$. 
This implies that $i = -1$ because $c_0$ is the only non-zero
value in $c$. But in that case, we have $c'_0 = 0$, and as
$\lLA$ can not return more than $1$, $c_0 = 2$ cannot be reached.
This is a contradiction.
\end{proof}

\begin{proof}[of Proposition~\ref{prop:X}]
Consider the two periodic configurations $c$ and
$c'$ defined as follows (see Figure~\ref{fig:X}):
$$\forall i \in \Z, \qquad \left\{\begin{array}{l}
    c_{2i} = 0\\
    c_{2i+1} = 1\enspace,
  \end{array}\right.\quad
\left\{\begin{array}{l}
    c'_{2i} = 0\\
    c'_{2i+1} = 2\enspace.
  \end{array}\right.
$$
It can be easily verified that $\fX(c) = \fX(c') = c$. Hence \X is
not \P-injective and, of course, it is not injective.

Let us prove that \X is \F-injective. Let $c$ and $c'$ be two distinct
finite configurations, and suppose that their image by $\fX$ is
identical. As the two configurations are finite, we can define $i \in
\Z$ being the least integer such that $c_i \ne c'_i$. As $\lX$ returns
only $0$ or $-1$, we know that $|c_i - c'_i| = 1$, and we can suppose
that $c_i = c'_i + 1$. That means that the local rule applied to $c$
at position $i$ is one of the seven rules which return $-1$:
\begin{itemize}
\item if the neighborhood is $(+\infty, -, -, -)$ (to make the
  notations clearer, $-$ represents any value), then since $c'_i
  = c_i-1$ and $c'_{i-2} = c_{i-2}$, the same rule is applied to
  $c'$, which means that $\fX(c)_i \ne \fX(c')_i$ which is a
  contradiction;
\item if the neighborhood is $(2, -, -, -)$, for the same reason the
  rule for the neighborhood $(+\infty, -, -, -)$ is applied to $c'$,
  which raises the same contradiction;
\item again, if the neighborhood is $(1, -1, -, -)$, $(1, -2, -, -)$
  or $(1, -\infty, -, -)$, the rule for the neighborhood $(2, -, -,
  -)$ is applied to $c'$, making $c'_i$ decrease by $1$;
\item if the neighborhood is $(0, -2, -, -)$ or $(0, -\infty, -,
  -)$, because $c'_{i-2} = c_{i-2}$, $c'_{i-1} = c_{i-1}$ and $c'_i
  = c_i-1$, one of the rules corresponding to the neighborhoods $(1,
  -1, -, -)$, $(1, -2, -, -)$ or $(1, -\infty, -, -)$ is applied to
  $c'$. There again, we have $\fX(c)_i \ne \fX(c')_i$\enspace.
\end{itemize}
\end{proof}

\begin{proof}[of Proposition~\ref{prop:surjp_surj}]
  Nearly exactly the same proof as for
  Proposition~\ref{prop:surjf_surj} can be made. The only change is
  that it starts with $c^0_n$ defined as the $(2n+1, \ldots,
  2n+1)$-periodic configuration with $\forall i\in\Z^d, |i| \leq n,
  \:(c^0_n)_i=c_i$.
  Everything else is unchanged.
\end{proof}

\noindent\textbf{Proofs of Section~\ref{sec:basicsetp}.}\\[1ex]
\begin{proof}[of Corollary~\ref{cor:surjf-surjp}]
  \F-surjectivity implies surjectivity
  (Proposition~\ref{prop:surjf_surj}), which implies in dimension 1
  \P-surjectivity (Proposition~\ref{prop:surj_surjp}).
\end{proof}
\end{document}